\documentclass[10pt]{iopart}
\usepackage[utf8]{inputenc}
\usepackage{subfig}
\usepackage{geometry}
\usepackage{tikz}
\usepackage{graphicx}
\usepackage{pgfplots}
\expandafter\let\csname equation*\endcsname\relax
\expandafter\let\csname endequation*\endcsname\relax
\usepackage{mathtools}
\usepackage{xcolor}
\usepackage{textcomp}
\DeclarePairedDelimiter\abs{\lvert}{\rvert}%

\usetikzlibrary{external}

\pgfplotsset{compat=newest}
\begin{document}
	\title{Design of a Head Coil for High Resolution Mouse Brain Perfusion Imaging using Magnetic Particle Imaging}
	
	\author{Matthias Graeser$^{1,2}$, Peter Ludewig$^{3}$, Patryk Szwargulski$^{1,2}$, Fynn Foerger$^{1,2}$, Tom Liebing$^{1,2}$, Nils D. Forkert$^{4}$, Florian Thieben$^{1,2}$, Tim Magnus$^{3}$, Tobias Knopp$^{1,2}$}
	
	\address{$^{1}$Section for Biomedical Imaging, Department of Diagnostic and Interventional Radiology and Nuclear Medicine at the University Medical Center Hamburg‐ Eppendorf, Hamburg, Germany}
	\address{$^{2}$Institute for Biomedical Imaging, Hamburg University of Technology, Hamburg, Germany}
	\address{$^{3}$Department of Neurology at the University Medical Center Hamburg‐ Eppendorf, Hamburg, Germany }
	\address{$^{4}$Department of Radiology and Hotchkiss Brain Institute, University of Calgary, Canada}
	\ead{ma.graeser@uke.de}    
	\def\iron{$_\textrm{Fe}$}
	\date{April 2019}
	\begin{abstract}
		Magnetic Particle Imaging (MPI) is a novel and versatile imaging modality developing towards human application. When up-scaling to human size, the sensitivity of the systems naturally drops as the coil sensitivity depends on the bore diameter. Thus, new methods to push the sensitivity limit further have to be investigated to cope for this loss. In this paper a dedicated surface coil for mice is developed, improving the sensitivity in cerebral imaging applications. Similar to MRI the developed surface coil improves the sensitivity due to the closer vicinity to the region of interest. With the developed surface coil presented in this work, it is possible to image tracer samples containing only 896\,pg\iron\;and detect even small vessels and anatomical structures within a wild type mouse model. As current sensitivity measures require a tracer system a new method for determining a sensitivity measure without this requirement is presented and verified to enable comparison between MPI receiver systems. 
	\end{abstract}
%	\pdfmapline{-dummy CMR10}
%\pdfmapline{-dummy CMR7}

\maketitle
\section{Preprint notice}
This is the version of the article before peer review or editing, as submitted by an author to IOP Physics in Medicine and Biology. IOP Publishing Ltd is not responsible for any errors or omissions in this version of the manuscript or any version derived from it. The Version of Record is available online at https://doi.org/10.1088/1361-6560/abc09e
\section{Introduction}

MPI is a novel and versatile tomographic imaging technology, which images the distribution of superparamagnetic iron oxide nanoparticles (SPIONs) \textit{in vivo} \cite{Gleich2005Nature}. Beside the determination of brain perfusion and stroke imaging\cite{ludewig2017magnetic}, MPI has proven to be suitable for a variety of medical applications like lung perfusion, gut bleeding, cardiovascular intervention and trauma detection \cite{Orendorff_2017,zhou2017first,haegele2016magnetic,YU2017}. Due to the sensitivity of the particles to their molecular micro-environment, MPI is also able to image properties like viscosity\cite{M_ddel_2018,Utkur2019}, temperature \cite{stehning2016simultaneous} and binding state \cite{VIERECK2017}. 
After rapid development and medical application experiments of MPI, the technology develops towards human application \cite{rahmer2017interactive, Graeser2019}. However, besides technical obstacles like, power requirements and electrical safety regulations, several physical and physiological issues like peripheral nerve stimulation arise.  As a physical consequence of up scaling the sensitivity of human systems decrease in comparison to pre-clinical systems \cite{Weizenecker2009PhysMedBio,Graeser2017SR}. Thus, new concepts for the receiver techniques have to be developed. One possibility to enhance the systems sensitivity is to use dedicated surface coils for specific regions, comparably to magnetic resonance imaging (MRI). With this the signal quality is enhanced as the specialized coil is closer to the region of interest (ROI). In addition, strong signals from other parts of the body are reduced as the sensitivity profile of a surface coil can be designed to damp areas outside the region of interest. However, in contrast to MRI, MPI signals are broadband, leading to own issues when using non rigid surface coils. This change of the transfer function can lead, if uncorrected, to artifacts within the image or even to a total failure of the reconstruction. Second, the position of the coil within the scanner is not constant, which makes coil sensitivity correction and passive decoupling harder to achieve. However, the sensitivity change does not lead to a reconstruction fail, but makes quantitative results more challenging. 
 
In this work, we present a head surface receive coil designed to achieve a very high signal quality within the head of a living mouse. In addition, a measurement protocol to decouple sensitivity measurements from the MPI contrast agent is developed to compare the sensitivity enhancement to other devices. The coil design is based on the digital mouse atlas, digimouse \cite{Dogdas_2007,Stout2002} and scaled to fit the typical mouse size used in our research laboratory. To be able to cancel out remaining background a second coil is connected to the first building a first order gradiometer. The whole setup is mounted within an animal support unit, ensuring the life support of the animal and at the same time keeping the workflow easy and simple. Due to the small size of the coil the measurement based approach for system matrix (SM) determination is not feasible. Thus, a hybrid approach \cite{Gruettner2011,vonGladiss2017} was used by correcting the SM of another receive coil with its measured transfer function to match the signals of the mouse surface coil. To investigate the advantages of the surface coil, a detailed comparison between the developed brain specific coil and a larger, non-optimized body coil is performed.

In MPI the sensitivity of different systems is hard to compare, as it is also dependent on the tracer type and its stability. In order to overcome this problem the system performance was characterized using relative measures to make the results independent of the particle systems \cite{Wells2017, Paysen2018}. However, in order to calculate the relative values, the particle sample to acquire the data has to stay stable for all measurements that are part of the procedure. In order to compare systems at different sites this assumption might not hold true as particle systems are subject to strict storage conditions defined by the manufacturer which might not be guaranteed when shipping small samples. Other work refer to particle independent measures like the coil noise and coil sensitivity\cite{Weizenecker2009PhysMedBio}, however this only translates to a system sensitivity measure if the system is dominated by the coil noise. To overcome these limitations, we developed a protocol which requires no particle system and thus determines the sensitivity of the instrumentation alone. It includes all noise and background sources influencing the sensitivity. The goal is to calculate a frequency dependent equivalent magnetic moment of the systems noise level . With this method, different MPI receivers can be compared with an easy-to-calculate and reliable measure. Using this method the coils are compared first by calculating the equivalent magnetic moment of the noise level, second with an SNR analysis based on a system matrix and third in the reconstructed image space showing a detection level within the picogram regime. After the comparison, the improved imaging performance of the brain specific coil is used to determine the brain perfusion of a healthy mouse. Compared to former MPI studies, the spatial resolution is high enough to identify small anatomical structures within the brain, like the retina or the brain stem.

\section{Methods}
\subsection{Mouse Atlas + 3D Mouse Model}
To design an organ specific coil it is first necessary to determine the location of the organ and the outer shape of the local body surface. For this, a digital mouse atlas was used (Digimouse: 3D Mouse Atlas \cite{Dogdas_2007,Stout2002}). This dataset offers pre-segmented data for each organ which can be exported in stereolithography (STL) file format. The resulting STL model was scaled to fit the mean weight of the mice used in our facility. The resulting mouse model was printed using an fused deposition modeling 3D printer (Ultimaker 3, Ultimaker Ltd.). This allows us to test the handling of the structural parts without the necessity of living animals as suggested in \cite{exner2019}.

\begin{figure}
    \centering
    \includegraphics[width=\textwidth] {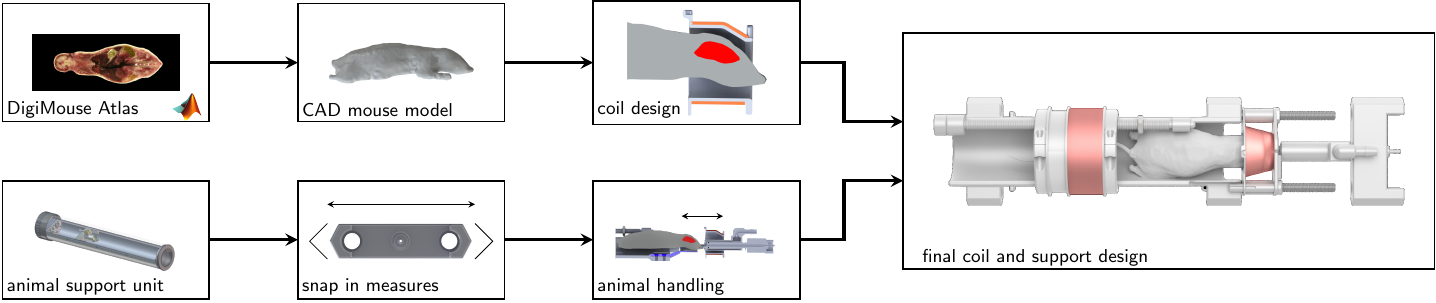}
    \caption{Design process. To achieve an easy-to-use setup with good receive properties the overall design of the system is influenced by two elements. On the one hand, the shape and position of the coil is designed based of the digimouse mouse atlas provided as Matlab (MathWorks) data files to maximize the sensitivity. On the other hand, the handling and fixation of the setup is mostly influenced by the animal support unit and the requirement for easy animal handling. Both requirements are taken into account resulting in the presented design.}
    \label{fig:Sketch}
\end{figure}

The mouse model was also imported in the computer-aided-design (CAD) model of the coil support to optimize the processes of animal handling and usability (see figure \ref{fig:Sketch}). As the system is aimed to be used by medical personal the handling in an \textit{in vivo} setting was reviewed and optimized together with an experienced neurologist. 

\subsection{Coil and Support Unit Design}
\begin{figure}[t]
     \centering
     \begin{tikzpicture}
     \node[xscale=1,yscale=1,name=model]{\includegraphics[width=0.85\textwidth]{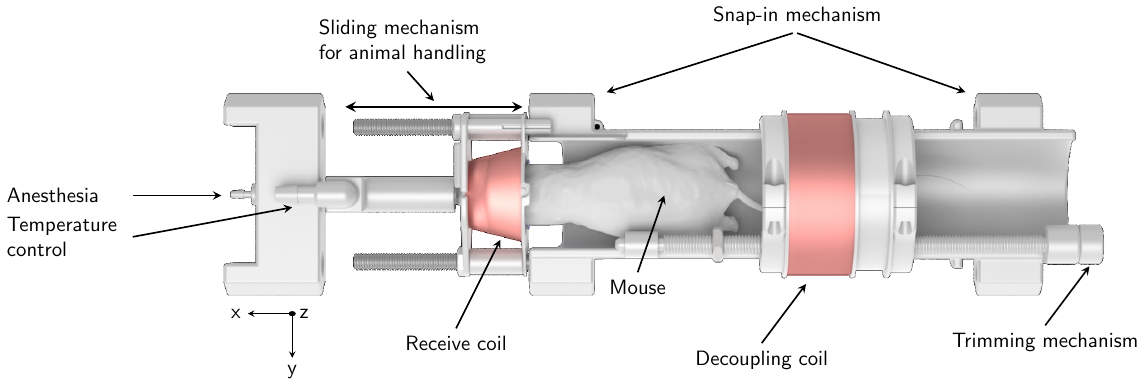}};
     \node[anchor=north] at ({model.south}) {\includegraphics[width=0.7\textwidth]{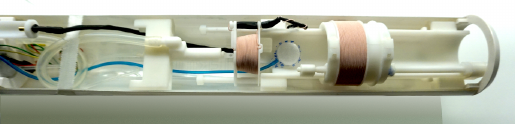}};
     \end{tikzpicture}

     \caption{Model and implementation of the coil support with the animal support unit. The receive coil is build as a first order gradiometer consisting of an anatomical shaped head coil and a solenoid cancellation coil. To place the mouse within the setup the receive coil is first pushed over the anaesthetic gas guidance. After the mouse is placed within the system the coil is pushed back over the head and fixated with two nuts.}
     \label{fig:ModelImplementation}
 \end{figure}

To ensure a good and reproducible fixture of the coil a specialized mouse bed was designed, which keeps the mouse in the center of the bore. As animal support unit a rat imaging cell (VO2 Rat imaging cell, Minerve, Esternay, France) is used. The mouse bed is designed in a way that it can be snapped in the support unit without the need of tools. The whole unit is then fixated at the end of the rat support to have a defined position. To ensure a reproducible position of the imaging cell within the scanner bore, a defined stop is added in the guidance system of the imaging cell. 
\\
 From the skin surface the minimum bore diameter necessary to fit the mouse head was determined. To optimize the coupling between the tracer and the receiver coil the coil diameter is reduced towards the nose region, matching the conical shape of the mouse head. As the signal intensity in the mouse brain is very weak compared to the heart, the length of the coil is chosen to be short to restrict the sensitive region to the brain and reduce the influence of the strong signals caused by the high tracer concentration in the heart and neck region. In total, the coil has a minimum width of 16.4\,mm and a minimum height of 17.4\,mm at the proximal end of the coil. 
 In the distal end of the coil the dimension widens to a maximum width of 25\,mm and a maximum height of 21\,mm. On the coil frame a total of 120 turns is wound. To attenuate unwanted background signals and the direct feed-through of the drive field, the receive coil is connected to a larger solenoid cancellation coil at the end of the mouse support in opposite turn direction (see figure \ref{fig:ModelImplementation}).

 The cancellation coil frame can be shifted along the proximal-distal axis by a screw driven sliding mechanism. Thus, the relative position for optimal background suppression can be adapted while mounted within the scanner. At the optimal position an attenuation of $>$60\,dB was achieved. To further dampen the remaining feed-though at the fundamental frequency a fourth order resonant filter with an additional attenuation of $>$ 90\,dB was inserted between receive coil and low noise amplifier. 
 
 An important feature of the device is an easy and simple animal handling. To position the mouse in a fast and secure way the receive coil can be loosened and slid over the anaesthesia unit. First an breath monitoring pad is placed on the torso region of the bed. After the mouse is placed securely on the support with its central tooth in the anaesthesia feeder, the receive coil is slid back over the head of the mouse and fixed with two nuts. To support the imaging cell within the scanner bore a second coil with a diameter of 72 mm was mounted as well, allowing simultaneous measurements with both coils.

\subsection{System Matrix Transfer-Function Correction}
\label{SM}
For Lissajous type systems as the MPI system used within this study (Bruker/Philips MPI Preclinical Imaging System) a system matrix reconstruction is the established reconstruction method \cite{Gruettner2011}.
As the minimal dimensions are 16.4 mm times 17.4 mm a measurement approach of the SM is not feasible. In addition, both sides of the bore are filled either by the robot or by the animal support making it impossible to provide visible feedback of the sample position. Thus, in this study a different approach was taken. As shown before \cite{graeser2013analog, vonGladiss2017,Halkola2013}, the phase and amplitude alternation by the amplifier and filter electronics can be recorded and corrected beforehand. Thus, a SM recorded using a 42\,mm receive coil was corrected by its transfer function to resemble the magnetic moment caused by the tracer. As tracer material Perimag (micromod Partikeltechnologie GmbH, Rostock) was used with a concentration of  8.5\,g\iron\;/ L$^{-1}$. The SM was recorded on a $ 25\times 25 \times 13$ grid with 1\,mm grid spacing and 100 averaged frames. As the signal quality suggests very high resolution, the system matrix was interpolated to a $50\times 50 \times 39$ grid resembling a pixel spacing of 0.5\,mm in $x$ and $y$ direction and 0.33\,mm in $z$ direction. Thus a reduction in spatial resolution due to a too coarse grid is prevented. In addition, the data of the phantom and \textit{in vivo} experiments were recorded using the head coil and were corrected by the transfer function of the head surface receiver. With both, the system matrix and the measurement in the domain of the magnetic moment, the reconstruction is made possible across different receive chains. This technique is analog to the hybrid approach presented in \cite{Gruettner2011,vonGladiss2017,Halkola2013} although the same system for measurement and SM determination is used.

\subsection{Sensitivity}
To determine the sensitivity two approaches were taken. First, the mean detection noise level in magnetic moment was calculated, second images of a dilution series were acquired following the protocol described in \cite{Graeser2017SR}.
\subsubsection{Equivalent Magnetic Moment Noise Level}
To determine the detection noise floor the following procedure was performed. First, an empty 3D measurement with a drive field amplitude set to zero is measured for 438.848\,s. Based on the spectral resolution of 46.4Hz and a bandwidth of 1.25MHz the measurement signal $\hat{u}(f)$ can be split in 20000 single frames containing 26929 frequencies per receive channel. We note that until now $\hat{u}(f)$ are integer values and not yet in the dimension of voltage. To match the measurement time of 2.14\, used in previous studies \cite{Graeser2017SR} a block averaging of $N=200$ frames was applied resulting in a measurement vector $\hat{u}^z_k(f)$. Here $k$ is the index of the averaged frames ($k=1,2,\ldots,100$),$f$ is the single frequency index ($f=1,2,\ldots,26929$) and $z$ is the available receive channel index ($z\in {x,y,z}$). In a second step, the magnitude of the $M=100$ remaining frames $\hat{u}^z_k(f)$ is calculated and averaged resulting in an equivalent noise level spectrum:
\begin{equation}
\hat{u}^z(f) = \frac{1}{M} \sum_{k=1}^{M} \abs{\hat{u}^z_k(f)}
\label{EQAbsAvg}
\end{equation}

Third, this noise floor is divided by the absolute value of the measured transfer function $TF(f)$. As the analog-digital-converter (ADC) reports not the received voltage, but corresponding integer values, it has to be corrected by an ADC specific calibration factor $c_\textrm{ADC}$ to achieve the equivalent magnetic moment (emm) of the noise floor.

\begin{equation}
    emm(f)=\frac{c_\textrm{ADC}\cdot u(f)}{\abs{\textrm{TF}(f)}}
\end{equation}
    	\pgfplotsset{xtick scale label code/.code={}}
	\pgfplotsset{ytick scale label code/.code={}}
The calibration factor can be acquired by applying a well known voltage to the ADC and recording the integer values recorded by the ADC. In this work a calibrated oscilloscope together with a signal source was used to apply a known sinusoidal signal to the ADC. After that the mean amplitude of the integer wave of a long measurement recorded by the ADC in time domain was taken. The ratio of the voltage applied in and the integer amplitude is the calibration factor $c_\textrm{ADC}$.

The presented procedure can also be applied with drive field on, which then includes the background of the system in the consideration of the sensitivity. To exclude the stable background which is typically subtracted equation \ref{EQAbsAvg} changes to:
\begin{equation}
\hat{u}^z(f) = \frac{1}{M-1} \sum_{k=2}^{M} \abs{\hat{u}^z_k(f)-\hat{u}^z_1(f)}
\end{equation}
\subsubsection{SNR Analysis}
Another way to analyze the system performance is the comparison of the signal-to-noise ratio (SNR) based on the system matrix \cite{Franke2016}. As it is not possible to measure the full field of view (FOV) at the given sequence parameters (12\,mT drive field, 2.5\,T\,m$^{-1}$ gradient) in the small receive coil, the SM was measured using the focus fields to emulate different sample positions \cite{vonGladiss2017, Halkola2013}. The sample is placed in the center, thus only the sensitivity in the center is included in the calibration. As the center of the coil has the lower sensitivity compared to the coil border, the SNR boost is underestimated slightly.
The calibration sample was chosen to be a delta sample size of $2\times2\times1$ mm$^3$ Perimag (micromod Partikeltechnologie GmbH, Rostock) with a concentration of 5\,mg\iron\;\,mL$^{-1}$(89\,mmol L$^{-1}$). The FOV was scanned at $14\times14\times14$ positions, with 20 averaged frames at 2.5\,T\,m$^{-1}$ gradient using a field based approach \cite{Halkola2013,vonGladiss2017}. To calculate the SNR two additional background SMs without tracer sample were used to subtract the background signals and calculate the mean noise and background drift. 

The SNR is then calculated by \begin{equation}
    \text{SNR}(f) = \frac{\left\Vert\mathbf{FG}(f)-\mathbf{BG}_1(f)\right\Vert_1}{\left\Vert(\mathbf{BG}_1 (f)-\mathbf{BG}_2(f)\right\Vert_1},
    \label{EQ:SNR}
\end{equation}
where $\mathbf{FG}(f)$ is the row of the foreground SM at frequency $f$ containing the tracer sample and $\mathbf{BG}_1(f)$ and $\mathbf{BG}_2(f)$ are the rows of two background SMs at frequency $f$. We use two full background matrices since the background signal is slightly focus-field dependent.
\subsubsection{Sensitivity in Image Space}
The sensitivity in image domain is performed by a dilution series using Perimag (micromod Partikeltechnologie GmbH, Rostock) as tracer material following the protocol developed in \cite{Graeser2017SR}. The total iron mass in the samples was varied from 459\,ng\iron\;down to 897\,pg\iron\;   in dilution steps of 2 resulting in 10 samples. All samples were moved along the $x$-axis to provide an image series showing a moving sample. A sample is defined as detected if the sample movement is resembled in the reconstructed images. All images were reconstructed using the system matrix pre processing described in \ref{SM}. As reconstruction technique a regularized Kaczmarz algorithim was applied using the reconstruction framework MPIReco \cite{Knopp2019MPIReco}.   
\subsection{\textit{In Vivo} Experiment Parameters}
All animal experiments were approved by local animal care committees (Behörde für Lebensmittelsicherheit und Veterinärwesen Hamburg, Nr. 16/41). We conducted the experiments following the “Guide for the Care and Use of Laboratory Animals” published by the US National Institutes of Health (NIH Publication No. 83-123, revised 1996), and performed all procedures in accordance with the ARRIVE guidelines (http://www.nc3rs.org/ARRIVE). C57BL/6 mice ( healthy animals: n = 3) were purchased from the Jackson Laboratory (Bar Harbor, ME 04609, USA). The mice were kept under a 12h light/dark cycle, at a constant temperature (22 $\pm$ 2$^\circ$C), and with food and water ad libitum. For imaging,  the mice were anesthetized with 1.5\% isoflurane (Abbott, Wiesbaden, Germany) in 100\% O2, and an intraperitoneal injection of 0.05 mg/kg body weight buprenorphine in saline.  The vital parameters were monitored using an animal support unit (Minerve, Esternay, France). For the tracer injection, a catheter (inner tube diameter 0.28 mm, Portex, Smiths Medical International Ltd., USA) was placed into the tail vein of the mouse.  We injected a total of 170\,\textmu g \iron\;of Perimag within a single bolus of 20\,\textmu L. As before, the gradient strength was set to2.5\,T\,m$^{-1}$ and the drive field amplitude to 12\,mT. The reconstruction were performed without any averaging providing the full spatial resolution of 46 frames per second. As a frequency selection a SNR threshold of 2.5 width a bandwidth of 40\,kHz to 450\,kHz was applied to the system matrix described in \ref{SM}. Only the x-channel was chosen for reconstruction. The regularization parameter of the Tikonov regularization within the reconstruction was chosen to 0.01. The same parameters were applied in the reconstruction of the body coil. However, we point out that the number of frequencies filtered by the SNR threshold are lower due to the lower performance of the body coil. The regularization factor was chosen to 0.1 in this case as for lower regularization factors the reconstruction fails.

\subsection{MRI Measurements and Post-Processing}
MRI images were obtained with a 7-Tesla small animal MRI (Bruker ClinScan, Billerica, MA, USA) using a dedicated transmit/receive (Bruker, Billerica, MA, USA) mouse body volume coil and a 3D gradient echo method with the following measurement parameters: TE=0.84\, ms, TR=10\, ms, flip angle=10$^\circ$, matrix 192 × 192 × 192 with elliptical k-space sampling, FOV=$32 \times 32 \times 32.64$ mm$^3$, 2 averages, readout bandwidth=700\,Hz per pixel and strong asymmetric echo readout.
Arterial vessels were measured using the time-of-flight (TOF) magnetic resonance angiography with the following scanning parameters: TE=3.06\$,ms, TR=17\,ms, flip angle=30$^\circ$, matrix size =256 × 256 × 24 with 16.7 percent slab oversampling and elliptical k-space sampling, FOV$\;=20 \times 20\,\textrm{mm}^2$, slice thickness 130\,\textmu m, 151\,Hz per pixel readout bandwidth and moderate asymmetric echo readout, a parallel acceleration factor of 2 with 19 inline acquired reference lines.
Perfusion analysis of the MPI datasets was performed using the software tool AnToNIa \cite{ForkertANTONIA2014}. Briefly described, the first three timepoints of the 4D MPI datasets were used to compute an average baseline image, which was then subtracted from each timepoint of the MPI scan to achieve a baseline intensity correction. After this, a b-spline approximation in the temporal dimension was used for smoothing of the signal intensity curve of each voxel in the MPI dataset. Finally, perfusion parameter maps of mean transit time (MTT), relative cerebral blood flow (rCBF), relative cerebral blood volume (rCBV), and time to maximum of the residual function (Tmax) were computed for each MPI dataset using block circulant single value decomposition employing a threshold of 0.15, whereas the arterial input function required for this was manually selected from the heart.

\section{Results}
\subsection{Sensitivity}
\subsubsection{Equivalent Magnetic Moment Noise Level}
\begin{figure}
    \centering
  \includegraphics[width=\textwidth]{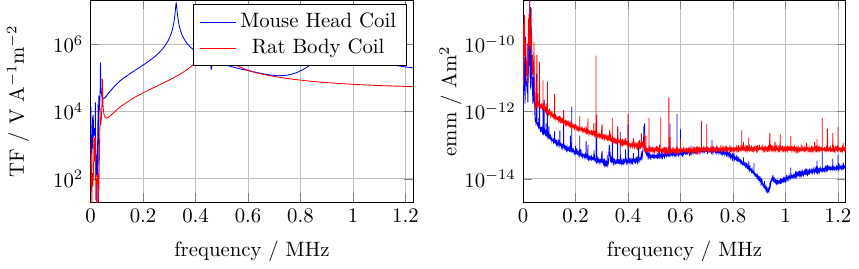}
    \caption{Transfer function (left) and equivalent magnetic moment of the noise level (emm) (right) for the 72\, mm coil (red) and the mouse head coil (blue). As both coils couple, the head coil shows a small ditch in at the resonance frequency of the 72\, mm coil. As signals from electronics within the shielding cabin can couple in the receive chain even without drive fields on, the emm shows various peaks above the noise level. However, there are only few peaks within the 26929 frequency components. The emm reveals a factor of $\approx 6 - 8$ in sensitivity gain by the specialized head coil, depending on the frequency range. }
    \label{Fig:TF_Noise}
\end{figure}

 Figure \ref{Fig:TF_Noise} (right) shows the equivalent magnetic moment (emm) of the noise spectrum caused by the receiver for a measurement time of 2.14\,s (N=200 complex valued averages). As can be seen, the minimum detection level is in the range of 300\,kHz, featuring a equivalent magnetic moment of 30\, fAm$^2$ at 320\,kHz. Due to the non-linear transfer function, the lowest detection level cannot be kept for a large frequency range. After the first resonance the impedance of the coil is larger than the impedance of the LNA resulting in a voltage drop before amplification. The second order resonance at 800\, kHz results in a higher effective amplification, and thus again reduces the noise equivalent magnetic moment. Compared to the 72\,mm coil, the sensitivity at 320\,kHz is an improvement by a factor of $\approx 6$. The most sensitive frequency region of the larger body coil is reached at around 600\, kHz. Here both coils feature approximately the same sensitivity. However, when measuring low concentrations of tracer, the bandwidth of the signal is limited to the lower region of the frequency band. Thus, thinking of \textit{in vivo} measurements the expected improvement is more in the range of 6. It is noted that the decrease of the emm in the lower frequency region is caused mainly by the inductive sensing which scales with the angular velocity $\omega$ thus, reducing the emm for high frequencies. 
\color{black}
\subsection{SNR Analysis}
\begin{figure}
    \centering
  \includegraphics[width=\textwidth]{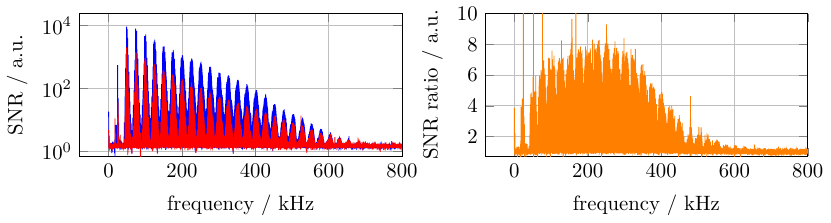}
    \caption{Results of the SNR measurement for the head coil (blue) amd the 72\, mm coil (red). The data is recorded using a 4\,\textmu L sample of Perimag with a concentration of 178.5\,mmolL$^{-1}$. The SNR values calculated by equation \ref{EQ:SNR} visualize the improvement in SNR by the brain specific surface coil. The right hand side shows the ratio between both SNR values, confirming the factor of 6-8 in the frequency range between 100\, kHz and 300\, kHz which was expected from the results in figure \ref{Fig:TF_Noise}.}
    \label{Fig:SNR}
\end{figure}

Figure \ref{Fig:SNR} shows the result of the SNR calculations based on the field based measured SMs. It can be seen that the brain specific surface coil outperforms the 72\, mm body coil. At around 600\, kHz which is for both receivers above their resonance frequency, both SNR values converge to the same level. In addition the ratio of the two SNR values was calculated revealing a SNR boost of around 6 to 8 in the most important frequency range for high sensitivity measurements. This confirms the expectation based on the detection level determined by the emm.  
\subsection{Dilution Series}
\begin{figure}
        \centering
   \includegraphics[width=\textwidth]{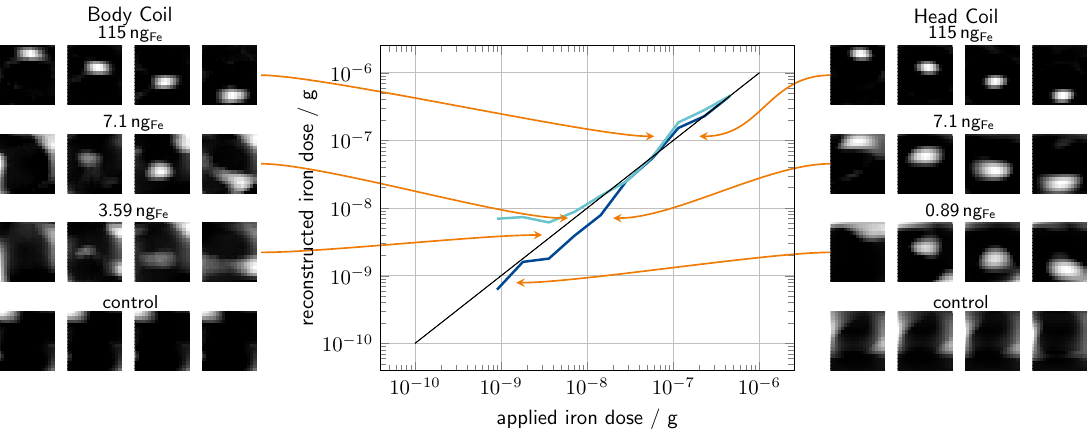}
    \caption{In vitro sensitivity study. A dilution series within the range of 459\,ng\iron\;to 896\,pg\iron\;was measured with the mouse head coil. The sample moved through the receiver coil along the $x$-axis. The sample is defined as detected if the movement of the sample could be observed within the image. The reconstructed iron concentraction was determined by summing the grey values in a region of interest around the sample position. The detection limit with the mouse head coil was at 897\,pg\iron\;while the detection limit for the body coil was 7.1\,ng\iron\;. Although the grey values in sum to a corresponding value at 3.59\,ng\iron\;for the body coil, the image reveals that the detection failed.}
    \label{Fig:Dilution}
\end{figure}

Figure \ref{Fig:Dilution} shows the results of the dilution series for the head coil. From the data of 10 different samples the 115\,ng\iron\;sample and the 896\,pg\iron\;sample are shown together with an empty robot drive as control. All 10 samples were successfully reconstructed and the sample movement was correctly identified. Thus, the detection limit in terms of iron mass is at least 896\,pg\iron\;for the tracer Perimag. 
 
\begin{figure}[ht]
	\includegraphics[width=\textwidth]{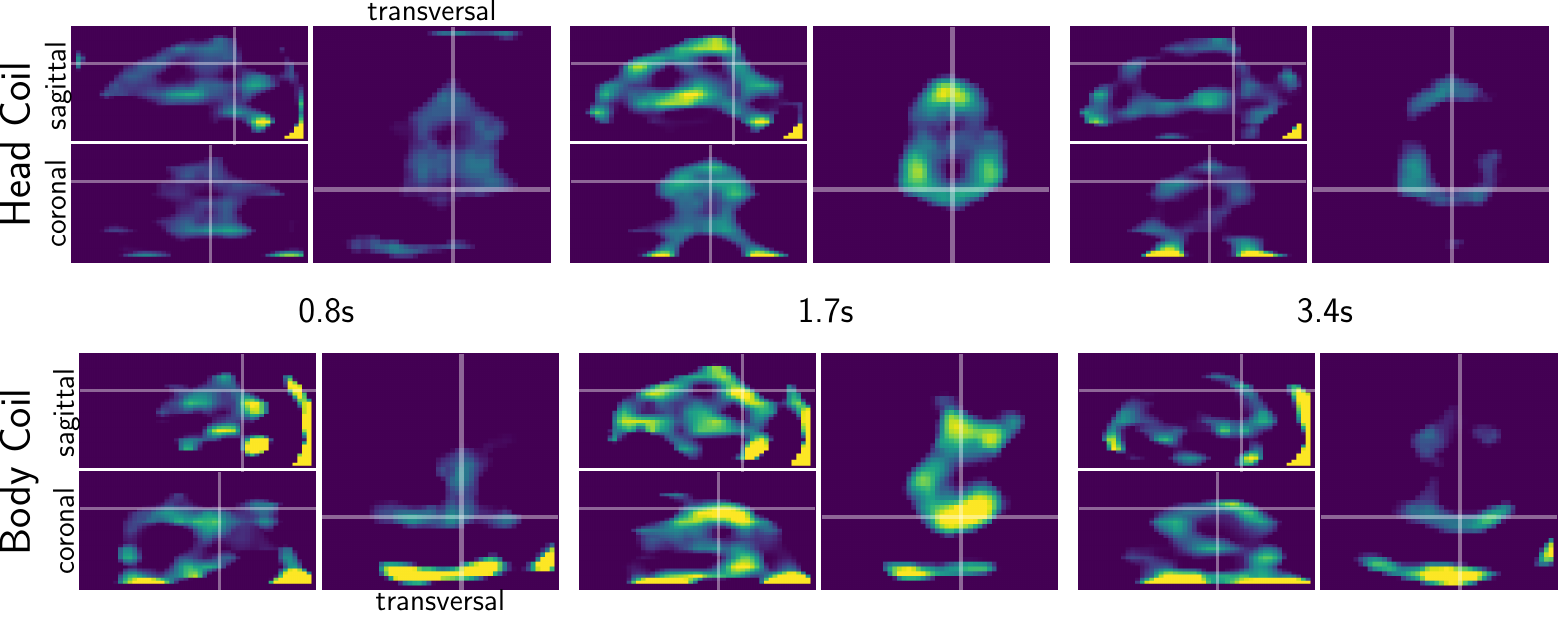}
	
	\caption{Timeseries of the \textit{in vivo} experiment for selected planes for the head coil (upper row) and the body coil (lower row). The time resolution for these reconstructed images is 21.14\,ms. Frame 40, 80 and 160 are shown. In the images the arterial perfusion of the mouse brain is seen at frames 40 to 80 (0.8\,s to 1.7\,s). The bolus then enters the venous system and is beginning to be flushed from the brain tissue at frame 160 (3.4\,s).
		The lower row shows the same data set for the body coil. The anatomical structures can only be detected when the peak of the bolus is within the brain. The image quality is much worse over the full time series. In addition, the strong influence of the heart area can be seen at the lower border of the transversal slice.}
	\label{fig:RecoMouse}
\end{figure}

\begin{figure}[h!]
	\centering
	\includegraphics[width=0.9\textwidth]{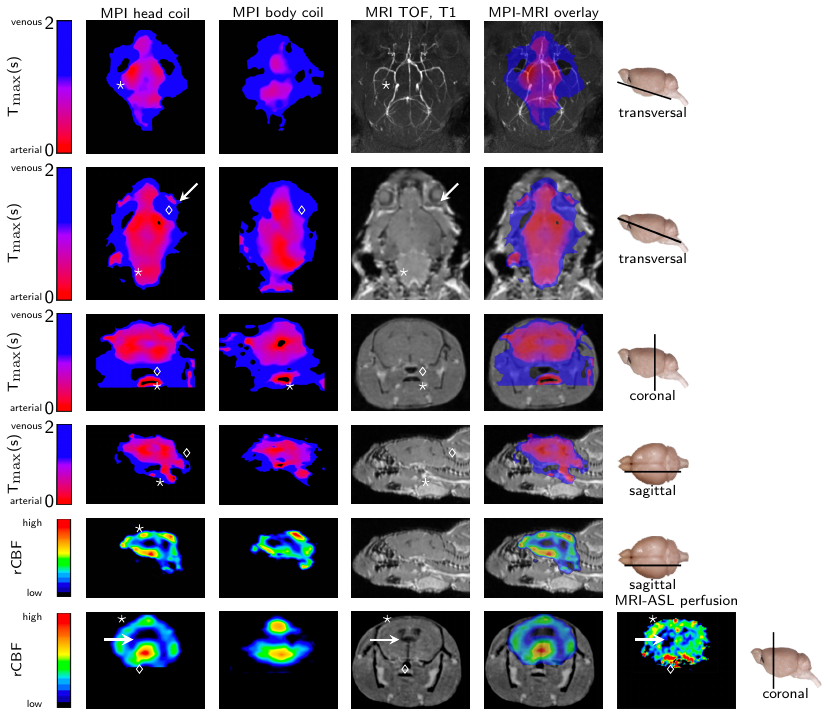}
	\caption{\textit{In vivo} comparison of MPI head vs. MPI body coil: Time-to-max parameter maps (Tmax, a-d) allow differentiation of arterial and venous signals and identification of anatomical structures in the brain. An arterial signal from the  circle of Willis (row 1, column 1, star) can be detected in the head coil data similar to MRI data (row 1, column 3, TOF: time-of flight angiography), whereas  this is not possible with data obtained with the body coil (column 2). The superior temporal and spatial resolution allows identification of small structures, like the retina (row 2, column 1/3, arrow), the retroorbital venous plexus (row 2, column 1, diamond), the brainstem (row 2, column 1/3, star), the nasopharynx (row 3, column 1/3, diamond), and the oral cavity (row 3, column 1-3, star). With the body coil data, it is not possible to detect retinal perfusion or to differentiate the brainstem and the nasopharynx. Blood supply from the anterior circulation via the internal carotid arteries and the posterior circulation via the vertebral arteries can be distinguished with the head coil (row 4, column 1/3, star: anterior circulation, diamond: posterior circulation). Comparison of the rCBF data (relative cerebral blood flow) from MRI (row 5, column 5) and MPI (row 5/6, column 1) shows similar results with increased blood flow in the cortex (row 5/6, column 1/3, star) and in the region of the circle of Willis (row 6, colum 1/3, diamond). Reduced blood flow is found in the striatum and around the ventricles (row 6, column 1/3/5, arrow). Due to lower sensitivity, the body coil rarely detects blood flow in the cerebral cortex (row 5/6, column 2). }
	\label{fig:PerfusionMaps}
\end{figure}

\subsection{\textit{In Vivo} Images: MPI Measurement of Cerebral Perfusion in C57Bl/6 Mice}
To compare the imaging quality of the head and body coils, we analyzed the cerebral perfusion in healthy mice. After one 20\,\textmu L single bolus of the MPI tracer Perimag (micromod Partikeltechnologie GmbH, Rostock), 3D real-time MPI data sets were obtained during the passage of the tracer through the brain in the full temporal resolution of 21.5\,ms (figure \ref{fig:RecoMouse}). Data sets were acquired simultaneously with both the body and head coil. 
While it was not possible to detect structures in the data set of the body coil, we could detect the anatomical structures of the brain in the head coil images without averaging. Due to the superimposition of arterial and venous MPI signals, it was challenging to distinguish anatomical structures.
To work around this problem, we calculated time-to-max (Tmax) parameter maps with the perfusion software AnToNIa (figure \ref{fig:PerfusionMaps}), which facilitate the distinction of arterial and venous vessels.
Due to the high sensitivity of the head coil, we could post-process the data with high temporal resolution and created very accurate parameter maps. With these high spatially and temporally resolved parameter maps, we detected fine structures in the mouse brain (figure \ref{fig:PerfusionMaps}, column 1), which was not possible with the previous body coil (figure \ref{fig:PerfusionMaps}, column 2). With this we could prove, that MPI is capable to visualize signals from the arteries of the circle of Willis (figure \ref{fig:PerfusionMaps}, row 1) and could differentiate between vessels of the anterior and posterior circulation (figure \ref{fig:PerfusionMaps}, row 3). We could even detect perfusion in the murine retina with MPI (figure \ref{fig:PerfusionMaps}, row 1-2) and could detect an increase in rCBF in the cerebral cortex of mice similar to MRI data (figure \ref{fig:PerfusionMaps}, row 5-6).

\section{Discussion}
In this paper, we introduce the use of specialized surface coils to maximize sensitivity in MPI.  In previous MPI studies, we had to make significant sacrifices regarding spatial or temporal resolution. Due to the low sensitivity of the body coil in our previous stroke study \cite{ludewig2017magnetic}, we could only achieve a spatial resolution of $3\times 3\times 1.5\textrm{mm}^3$ and needed to average the temporal resolution ten times to obtain usable images.  The hardware optimizations presented in this paper provide better spatial resolution at full temporal resolution (46 frames/s), which shows the possible potential of MPI.
With the adaptations it is now possible to detect perfusion in small arterial vessels in the brain and retina with a diameter of 150\,\textmu m.
With the spatial resolution of $0.50\times 0.50\times 0.39~\textrm{mm}^3$ at the full temporal resolution of 21.5\,ms, interesting scientific questions, e.g., with cell tracking, functional, and perfusion imaging, can now be addressed with MPI.
The presented coil shows a superior sensitivity in the picogram regime using Perimag marking a factor of 5.5 in comparison to previous published images. On the other hand, the tracer LS008, used in the study of Graeser et al. \cite{Graeser2017SR}, was two times more sensitive compared to Perimag, but was not available for our study. Thus, the improvement can be calculated to be in the range of one magnitude. 

We note that tracer systems are subject of aging and thus the performance can only be guaranteed in tolerances over finite time ranges. This reveals the necessity to develop methods for determining the sensitivity without the dependence on a tracer system. To make the comparison between different MPI systems independent from tracer measurements, we developed a method to quantify the sensitivity solely by the equivalent magnetic moment of the systems noise. A big advantage is the independence of the method from the used excitation sequence, which makes also comparison between systems with different encoding schemes possible. As the method can be performed with standard measurement technology for electronic labs, it allows the usage in all hardware oriented research groups. 

At typical pre-clinical dosage above 100\,\textmu g\iron\;in a pre-clinical scenario the image resolution and quality shows a large improvement compared to former results \cite{ludewig2017magnetic}. With the improved spatial resolution the perfusion information can be judged without the need of a cross-reference image like MRI. 
The results shown here demonstrate the potential of high sensitivity MPI systems. However, even these results probably do not provide the best image for the data recorded. In frequency space reconstruction, the resolution is limited to the voxel size of the system matrix. Reducing this further would reduce the signal to noise ratio by a power of three. With decreasing voxel size the SM becomes noise dominant in the frequency components providing the high resolution information. Thus, alternative ways to acquire the SM have to be taken. The hybrid approach using field based methods will allow the decoupling from voxel size to tracer mass allowing the measurement of high system matrix resolutions with high SNR \cite{vonGladiss2017,Graeser2017PMB}. Using such a high resolved SM the image resolution is expected to be below our current SM resolution.

In this first attempt a rigid coil based on the anatomy of a mouse head was designed. To design optimized coils for other organs in the torso region, such rigid bore coils are not the best choice. As known from MRI surface coils should adapt to the outer body surface. As the signal nature is broadband in MPI, this introduces its own problems as the transfer function is dependent on the inductance of the coil, which itself depends on the geometry of the coil. In addition, the coupling between drive field and receive coil is not fixed due to the unpredictable geometry and location of those flexible surface coils. To use flexible coil systems both issues have to be addressed. To correct the altered transfer function a correction term has to be determined. An easy way to achieve this is to include a small calibration coil winding within the flexible coil support. After adjustment, the transfer function could be recorded. This would already allow for qualitative imaging. For quantitative results, the coil sensitivity has to be included as well. As the geometry is unknown, this could be the most challenging task. A possible solution might be to access the geometry by photographic image, rigid deformation and Biot-Savart simulation of the new coil sensitivity profile. For this a suitable pattern could be added on the coil surface. 
To address the second issue, active suppression techniques could be implemented as presented before \cite{Zheng2013}. This should at least partly cope for the missing passive decoupling. 

The reached iron mass of 896\,pg\iron\;corresponds to approximately 69 mesenchymal stem cells if each take up 13\,pg\iron\;\cite{bulte2015quantitative}. This value can further be reduced by three approaches. The reconstructed frame in the dilution experiment had a measurement time of 2.14\,s. For stem cell tracking lower frame rates might be feasible leaving room for SNR improvements by further averaging. E.g. reducing the time resolution by a factor of 9 to 19.26\,s the number of cells drops by a factor of $\sqrt{9}$ to 23. A further improvement might be achieved by reducing the noise and the bandwidth of the receiver further with additional parallelisation of the low noise amplifier input stage. However, this would be at cost of image resolution for higher iron dosages due to the drop in bandwidth. Nevertheless, we expect another factor of 1.5 to be feasible in signal improvement using this approach. As noted before, there is at least a factor of 2 in sensitivity gain possible at the tracer signal performance. This was already demonstrated by the tracer system LS008 \cite{Khandhar2015,Ferguson2015MPILS008}. The most promising factor might be the uptake of iron within stem cells. As the tested tracers in \cite{bulte2015quantitative} were not optimized for cell uptake there might be potential to include more tracer material in a single cell. The potential is even more prominent by looking at the large variation in iron load of the different tracer materials. In combination this could result in a detection limit of less than 10 cells with the potential of even detecting single cells.

\section {Acknowledgments}
This work was supported by the “Forschungszentrums Medizintechnik Hamburg” (FMTHH) by the Hertie-Stiftung (Hertie Academy of Clinical Neuroscience), the German Research Foundation (DFG; grant numbers: GR 5287/2-1, KN 1108/7-1, DFG FOR 2879 [project LU 1924/1-1]). This work was also supported by the BMBF under the frame of EuroNanoMed III {(grant number: 13XP5060B, "Magnetise")}.
\section{Author Contributions}
M.G., P.S., P.L. and T.K. had the idea for the project. M.G. P.L. P.S. and T.L. worked on the design of the receive coil structure. T.L. constructed the coil and the support. M.G. F.F. and F.T. build the necessary receive electronics. P.L., P.S., T.L. and M.G. performed the measurements. P.S. T.L. and M.G. reconstructed the images. M.G. developed the noise based sensitivity measure. N.D.F., P.L. and T.M calculated the perfusion parameter maps. M.G. wrote the manuscript. All authors reviewed the paper. 
\section{References}
%\bibliography{ref}

\end{document}